\documentstyle [12pt] {article}
\title{COMPLETE QUANTUM THERMODYNAMICS OF THE BLACK BODY PHOTON GAS}

\author{Vladan Pankovi\'c, Darko V. Kapor\\
Department of Physics, Faculty of Sciences, 21000 Novi Sad,\\ Trg
Dositeja Obradovi\'ca 4, Serbia, \\vladan.pankovic@df.uns.ac.rs}

\date {}
\begin {document}
\maketitle \vspace {0.5cm}

 keywords: photon gas thermodynamics, black body, black hole
 entropy
\vspace {0.3cm}

\begin {abstract}
Kelly and Leff demonstrated and discussed formal and conceptual
similarities between basic thermodynamic formulas for the
classical ideal gas and black body photon gas. Leff pointed out
that thermodynamic formulas for the photon gas cannot be deduced
completely by thermodynamic methods since these formulas hold two
characteristic parameters, {\it r} and {\it b}, whose accurate
values can be obtained exclusively by accurate methods of the
quantum statistics (by explicit use of the Planck's  or
Bose-Einstein distribution). In this work we prove that the
complete quantum thermodynamics of the black body photon gas can
be done by simple, thermodynamic (non-statistical) methods. We
prove that both mentioned parameters and corresponding variables
(photons number and pressure) can be obtained very simply and
practically exactly (with relative error about few percent), by
non-statistical (without any use of the Planck's or Bose-Einstein
distribution), quantum thermodynamic methods. Corner-stone of
these methods represents a quantum thermodynamic stability
condition that is, in some degree, very similar to quantum
stability condition in the Bohr quantum atomic theory (de
Broglie's interpretation of the Bohr quantization postulate).
Finally, we discuss conceptual similarities between black body
photon gas entropy and Bekenstein-Hawking black hole entropy.
\end {abstract}

\vspace{1cm}

Kelly [1] and Leff [2] demonstrated and discussed formal and
conceptual similarities between basic thermodynamic formulas for
the classical ideal gas and black body photon gas. Leff pointed
out that thermodynamic formulas for the photon gas cannot be
deduced completely by thermodynamic methods since these formulas
hold two characteristic parameters, {\it r} and {\it b}, whose
accurate values can be obtained exclusively by accurate methods of
the quantum statistics (by explicit use of the Planck's  or
Bose-Einstein distribution). In this work we shall prove that the
complete quantum thermodynamics of the black body photon gas can
be done by simple, thermodynamic (non-statistical) methods.
Precisely, we shall prove that both mentioned parameters and
corresponding variables (photon number and pressure) can be
obtained very simply and practically exactly (with relative error
about few percent), by non-statistical (without any use of the
Planck's or Bose-Einstein distribution), quantum thermodynamic
methods. Corner-stone of these methods represents a quantum
thermodynamic stability condition which is, in some degree, very
similar to quantum stability condition in the Bohr quantum atomic
theory (de Broglie's interpretation of the Bohr quantization
postulate). Finally, we discuss conceptual similarities between
black body photon gas entropy and Bekenstein-Hawking black hole
entropy.

As it is well-known [3]-[5] quantum statistics of the black body
photon gas predicts accurately, using Planck's or Bose-Einstein
distribution, the following expressions for black body photon
numbers - $N$, internal energy - $U$, pressure - $p$ and entropy -
$S$
\begin {equation}
   N = \frac {16 \pi k^{3} \zeta (3)}{h^{3}c^{3}} T^{3} V = {\it r}V T^{3}
\end {equation}
\begin {equation}
   U = \frac {8\pi^{5}k^{4}}{15 h^{3}c^{3}} T^{4}V N = {\it b}V T^{4} = 2.7 N k T
\end {equation}
\begin {equation}
   p = \frac {1}{3}\frac {U}{V} = \frac {8\pi^{5}k^{4}}{45 h^{3}c^{3}} T^{4}= \frac {1}{3}{\it b}T^{4} = 0.9 \frac {N k T}{V}
\end {equation}
\begin {equation}
   S = \frac {4}{3}\frac {U}{T} = \frac {32\pi^{5}k^{4}}{45 h^{3}c^{3}} T^{3}V = \frac {4}{3}{\it b}VT^{3} = 3.6 N k  .
\end {equation}
Here $k$ represents the Boltzmann constant, $h$ - Planck constant,
$c$ - speed of light, $T$ - black body temperature, $V$ - black
body volume and $\zeta (3) \simeq 1.202$ - Riemann zeta function,
while
\begin {equation}
    {\it r} = 60.4 (\frac {k}{hc})^{3}
\end {equation}
\begin {equation}
     {\it b} = \frac {8\pi^{5}k^{4}}{15 h^{3}c^{3}}
\end {equation}
represent parameters that can not be obtained by relatively
inaccurate thermodynamic methods but that can be obtained exactly
by accurate methods of the quantum statistics (with explicit use
of the Planck's or Bose-Einstein distribution).

However, suppose that photon gas is captured in a hollow,
spherical black body with relatively large radius $R$ at
temperature $T$.

Suppose that, in the thermodynamic equilibrium, photons within
hollow, spherical black body propagate, practically, over black
body internal surface. Precisely, suppose that these photons
propagate over a black body great internal circle with radius $R$
and circumference $2\pi R$, according to the following quantum
thermodynamic stability condition
\begin {equation}
    <m>cR= <n> \frac {h}{2\pi}          .
\end {equation}
Here $<n>$ represents the thermodynamically averaged quantum
number or (roughly speaking) quantum state and $<m>$ -
thermodynamically averaged value of the (single) photon mass in
this quantum state.

Expression (7) implies
\begin {equation}
      2\pi R = <n> \frac {h}{<m>c} = <n> <\lambda>
\end {equation}
or, correspondingly,
\begin {equation}
      R = <n> \frac {<\lambda>}{2\pi} = <n> < \lambda_{R} >         .
\end {equation}
Here $<\lambda>=\frac {h}{<m>c}$ represents the thermodynamic
average wavelength of the black body photon radiation. It
corresponds to the thermodynamically averaged frequency of the
black body photon radiation $<\nu>=\frac {c}{<\lambda>}$. Also,
$<\lambda_{R}>=\frac {<\lambda>}{2\pi}$ in (9) represents
corresponding thermodynamic average reduced wavelength of the
black body photon radiation.

In this way quantum thermodynamic stability conditions (7)-(9)
simply mean the following. By simple thermodynamic averaging
circumference of the black body $2\pi R$ holds, in the quantum
state $<n>$, corresponding quantum number of the photon
wavelengths $<\lambda>$. Obviously, this condition, in some
degree, is similar to quantum stability condition in the Bohr
quantum atomic theory (de Broglie's interpretation of the Bohr
quantization postulate). But of course, in the Bohr atomic model,
that refers on the pure but not on the thermodynamically mixed
quantum state, there are different electron circular quantum
orbits corresponding to the different quantum states. Vice versa,
here different quantum states, dependent of the temperature,
correspond to the same photon circular orbit over black body
internal surface.

Suppose, according to the Planck formula and (9),
\begin {equation}
   <\nu> = \frac {kT}{h}
\end {equation}
so that
\begin {equation}
    <\lambda> = \frac {hc}{kT}
\end {equation}
and
\begin {equation}
    <n> = \frac {R}{< \lambda_{R}>} = \frac {2\pi R kT}{hc}       .
\end {equation}
It can be observed that while $<\nu>$ (10) and $<\lambda>$ (11)
represent functions of the $T$ only, $<n>$ (12) is function of the
$T$ and $R$.

Further, expression (9) can be simply transformed in the following
expression
\begin {equation}
  \frac {4}{3} \pi R^{3} =   \frac {4}{3}<n>^{3} < \lambda_{R}>^{3}
\end {equation}
or, since $\frac {4}{3} \pi R^{3}$ represents the black body
sphere volume $V$, in
\begin {equation}
  V = \frac {4}{3} \pi <n>^{3} < \lambda_{R}>^{3}   .
\end {equation}
   It yields
\begin {equation}
      <n>^{3} = \frac {3}/{4\pi} \frac {V}{< \lambda_{R}>^{3} }
\end {equation}
or, according to the previously introduced expressions (5), (10),
(11) ,
\begin {equation}
   <n>^{3}= 6\pi^{2}(\frac {k}{hc})^{3}T^{3}V = 59.2 (\frac {k}{hc})^{3}VT^{3}    .
\end {equation}

Now, it can be observed that term $59.2 (\frac {k}{hc})^{3}$ in
(16) is practically (with relative error about 2$\%$) identical to
parameter ${\it r}$ (5). It admits statement that
\begin {equation}
  {\it r} = 6\pi^{2}(\frac {k}{hc})^{3}
\end {equation}
represents correctly, simply quantum thermodynamically,
non-statistically (without any use of the Planck's or
Bose-Einstein distribution) derived value of the parameter ${\it
r}$.

Then, right hand of (16) is practically identical to right hand of
(5) so that we can suppose that left hand of (16) is practically
identical to left hand of (5) too, i.e.
\begin {equation}
      N = <n>^{3}             .
\end {equation}

In this way we reproduced correctly, practically accurately (with relative error about 2%), simply quantum thermodynamically, non-statistically (without any use of the Planck's or Bose-Einstein distribution), expression for the number of the photons emitted by a (spherical) black body in the thermodynamic equilibrium.

Further, it can be simply quantum thermodynamically,
non-statistically (without any use of the Planck's or
Bose-Einstein distribution), supposed that power, $P$, of the
black body radiation per unit area, $A$, can be approximated by
the following expression
\begin {equation}
   \frac {P}{A} =  \frac {<F>c}{A }= \frac {<\nu>}{<\lambda_{R}>^{2}} kT
\end {equation}
where $<F>$ represents the thermodynamic average force of the back
reaction of the radiation at the unit area.

According to (10), (11) and since
\begin {equation}
  \frac {\pi^{3}}{30} = 1.033 \simeq 1
\end {equation}
it follows
\begin {equation}
   \frac {P}{A} \simeq \frac {\pi^{3}}{30\frac {<\nu>}{<\lambda_{R}>^{2}}} kT =
   \frac {c}{4}{\it b}T^{4}\equiv \sigma T^{4}
\end {equation}
representing Stefan-Boltzmann law derived simply
thermodynamically, non-statistically (without any use of the
Planck's or Bose-Einstein statistics), where, of course, $\sigma$
represents the Stefan-Boltzmann constant. It can be observed that
this expression is very close (with relative error about 3$\%$) to
the exactly, quantum statistically obtained Stefan-Boltzmann law
(by explicit use of the Planck's  or Bose-Einstein distribution).

Expression (19) implies the following simple quantum
thermodynamic, non-statistical (without any use of the Planck's or
Bose-Einstein distribution), expression for the pressure of the
photon radiation emitted by black body
\begin {equation}
  p_{em} = \frac {<F>}{A} = \frac {1}{<\lambda><\lambda_{R}>^{2} }kT= 4\pi^{2} \frac {kT}{<\lambda>^{3}}   .
\end {equation}
It, according to (10), (11), (16), can be simply transformed in
the following way
\begin {equation}
  p_{em}  =  4\pi^{2} \frac {kT}{<\lambda>^{3}} =
  \frac {4}{6} 6\pi^{2} (\frac {k}{hc})^{3}T^{3} kT = \frac {2}{3} \frac {NkT}{V}  .
\end {equation}

On the other hand, according to (9)-(11), simple quantum
thermodynamic, non-statistical (without any use of the Planck's or
Bose-Einstein distribution), expression for the pressure of the
reflected photon radiation in a hollow spherical black body equals
approximately
\begin {equation}
  p_{ref} = N \frac {\frac {mc^{2}}{R}}{4\pi R^{2}} =
  \frac {1}{3} \frac {NkT}{\frac {4}{3}\pi R^{3}} = \frac {1}{3} \frac { NkT}{V}.
\end {equation}

Then, total pressure of the black body photon gas obtained by
simple quantum thermodynamic, non-statistical (without any use of
the Planck's or Bose-Einstein distribution) methods equals
\begin {equation}
  p = p_{em} + p_{ref} = \frac { NkT}{V}     .
\end {equation}
This expression is very close (with relative error about 10%) to the exact expression (3) obtained by accurate quantum statistical methods (by explicit use of the Planck's or Bose-Einstein distribution). It can be observed that this expression is formally-numerically identical to the expression  for pressure of a classical ideal gas, but expression for pressure of a classical ideal gas does not hold at all expression for the pressure of the back reaction of the emitted photons.

It can be added that when photon gas pressure is determined by
simple quantum thermodynamic, non-statistical (without any use of
the Planck's or Bose-Einstein distribution) methods (25),
parameter {\it b}, according to (3), can be determined by these
methods too.

In this way it can be concluded that complete quantum
thermodynamics of the black body photon gas can be deduced by
simple, quantum thermodynamic, non-statistical (without any use of
the Planck's or Bose-Einstein distribution) methods, even if, of
course, complete quantum thermodynamics of the photon gas
represents an approximate theory of the accurate quantum
statistical theory of the photon gas (with explicit use of the
Planck's  or Bose-Einstein distribution) .

Consider, finally entropy of the black body photon gas according
to simple, quantum thermodynamic, non-statistical (without any use
of the Planck's or Bose-Einstein distribution) methods. It, in
fact, represents introduction of (25) in (4) which, according to
(16), yields
\begin {equation}
 S = 4 kN = k 24\pi^{2} (\frac {kT}{hc})^{3} V = k \frac {3}{\pi}  \frac {V}{< \lambda_{R}>^{3} }\simeq k  \frac {V}{< \lambda_{R}>^{3} }      .
\end {equation}
Accurate quantum statistical expression for the black body photon
gas entropy (4) can be transformed in
\begin {equation}
  S = 3.6 Nk =  \frac {32\pi^{5}k^{4}}{45 h^{3}c^{3} } T^{3} V =
  k (\frac {4\pi^{2}}{45} ) \frac {V}{< \lambda_{R}>^{3} }\simeq 0.88 k  \frac {V}{< \lambda_{R}>^{3} }    .
\end {equation}
In both cases, i.e. for (26), (27), entropy of the black body
photon gas is practically equivalent to the quotient of the black
body volume $V$ and "minimal quantum thermodynamic volume"
$<\lambda_{R}>^{3}$ multiplied by Boltzmann constant $k$. It
represents very interesting result.

All this is similar to the Bekenstein formula for the black hole
entropy
\begin {equation}
 S_{BH} = k \frac {A_{BH}}{(2L_{P})^{2}}= k \frac {A_{BH}}{(2\lambda_{RP})^{2}}        .
\end {equation}
Here $ A_{BH}$ represents the surface of the black hole which, for
a spherical, Schwarzschild black hole with horizon radius $R_{S}$,
equals $A_{BH}=4\pi R^{2}_{S}$. Also, here $(2L_{P})^{2}$
represents "elementary surface of the black hole" since $L_{P} =
(\frac {\hbar c}{G})^{\frac {1}{2}}= \lambda_{RP}$ represents the
Planck length equivalent to the reduced Compton wave length of a
particle with Planck mass, where $\hbar=\frac {h}{2\pi}$
represents the reduced Planck constant and $G$ - Newtonian
gravitational constant. As it is well-known this Planck length
represents the minimal length with the physical meaning in the
quantum field theories.

Since entropy represents an additive variable, expression (26) we
can transform in the following way
\begin {equation}
 S  \simeq k \frac {V}{< \lambda_{R}>^{3}} \simeq k \frac {V_{shell}}{< \lambda_{R}>^{3}}  + k \frac {V_{int}}{< \lambda_{R}>^{3}} \equiv S_{shell} + S_{int} =
 k \frac {A}{<\lambda_{R}>^{2}} + S_{int}   .
\end {equation}
Here
\begin {equation}
 S_{shell} = k \frac {V_{shell}}{< \lambda_{R}>^{3} }= k \frac {A<\lambda_{R}>}{< \lambda_{R}>^{3} }= k \frac {A}{<\lambda_{R}>^{2}}
\end {equation}
can be considered as the entropy in a tiny (with width
$<\lambda_{R}>$) spherical shell nearly black body surface $A=4\pi
R^{2}$, while $S_{int} = S - S_{shell}$ represents the entropy of
the rest, internal (under shell) volume of the black body photon
gas.

It can be observed that Bekenstein black hole entropy (28) has
practically identical form as the entropy of the black body photon
gas spherical shell (30). It is not any surprise since black hole
represents, at it has been proved by Bekenstein, Hawking and other
[6], [7], an especial form of the black body. Classically
speaking, according to the cosmic censorship conjuncture, except
information on the black hole total mass-energy (for a
Schwarzschild black hole), there is no other information on the
black hole physical characteristics inside horizon. Quantum field
theory, however, roughly speaking, according to the possibility of
the particle-antiparticle pair creation nearly horizon, admits
additional considerations of the events that occur in
$2\lambda_{RP}$ a tiny spherical shell (with width about
$2\lambda_{RP}$) nearly horizon corresponding to the
Bekenstein-Hawking black hole entropy as, practically, black body
shell entropy. On the other hand, according to the cosmic
censorship conjuncture, black hole does not hold internal (under
shell) entropy in difference to the other black bodies for which
cosmic censorship conjuncture is not satisfied.

In conclusion we can shortly repeat and point out the following.
Kelly and Leff demonstrated and discussed formal and conceptual
similarities between basic thermodynamic formulas for the
classical ideal gas and black body photon gas. Leff pointed out
that thermodynamic formulas for the photon gas cannot be deduced
completely by thermodynamic methods since these formulas hold two
characteristic parameters, {\it r} and {\it b}, whose accurate
values can be obtained exclusively by accurate methods of the
quantum statistics (by explicit use of the Planck's  or
Bose-Einstein distribution). In this work we prove that the
complete quantum thermodynamics of the black body photon gas can
be done by simple, thermodynamic (non-statistical) methods. We
prove that both mentioned parameters and corresponding variables
(photon number and pressure) can be obtained very simply and
practically exactly (with relative error about few percent), by
non-statistical (without any use of the Planck's or Bose-Einstein
distribution), quantum thermodynamic methods. Corner-stone of
these methods represents a quantum thermodynamic stability
condition that is, in some degree, very similar to quantum
stability condition in the Bohr quantum atomic theory (de
Broglie's interpretation of the Bohr quantization postulate).
Finally, we discuss conceptual similarities between black body
photon gas entropy and Bekenstein-Hawking black hole entropy.

\vspace{1cm}

{\large \bf References}

\begin{itemize}

\item [[1]]  R. Kelly, Am. J. Phys., {\bf 49} (1981) 714
\item [[2]] H. D. Leff, Am. J. Phys. {\bf 70} (2002) 792
\item [[3]] D. J. Amit, J. Werbin, {\it Statistical Physics: An Introductory Course} (World scientific, Singapore, 1995)
\item [[4]] L. D. Landau, E. M. Lifshitz, {\it Statistical Physics } (Adison-Wesley, Reading, MA, 1958)
\item [[5]] A. Isihara, {\it Statistical Physics } (Academic Press, New York, 1971)
\item [[6]] R. M. Wald, {\it Quantum Field Theory in Curved Spacetime and Black Hole Thermodynamics} (University of Chicago Press, Chicago, 1994)
\item [[7]] R. M. Wald, {\it The Thermodynamics of Black Holes}, gr-qc/9912119

\end {itemize}

\end {document}